\documentclass[
reprint,
amsmath,amssymb,
aps,
]{revtex4-1}

\usepackage{graphicx}
\usepackage{dcolumn}
\usepackage{bm}

\begin{document}

\preprint{}

\title{Emptiness Formation Probability in 1D Bose Liquids}

\author{Hsiu-Chung Yeh}
\affiliation{School of Physics and Astronomy, University of Minnesota, Minneapolis, Minnesota 55455, USA}

\author{Alex Kamenev}
\affiliation{School of Physics and Astronomy, University of Minnesota, Minneapolis, Minnesota 55455, USA}
\affiliation{William I. Fine Theoretical Physics Institute, University of Minnesota, Minneapolis, Minnesota 55455, USA}

\begin{abstract}
We study  emptiness formation probability (EFP) in interacting 1D Bose liquids. That is the probability 
that a snapshot of its ground state reveals exactly zero number of particles within the interval $|x|<R$. For a weakly 
interacting liquid there is parametrically wide regime $n^{-1} < R <\xi$ (here $n$ is the average density and $\xi$ is the healing length), where EFP exhibits a non-trivial crossover from the Poisson to the Gaussian behavior. We employ the instanton technique [A. Abanov, 2004] to study quantitative details of these regime and compare it with previously reported limited cases.
\end{abstract}

\maketitle


\section{Introduction}

Recent  precision measurements of particle number fluctuations in ultra cold quantum gases \cite{esteve2006observations,armijo2010probing,jacqmin2011sub}
have revived interest \cite{del2011long,pons2012fidelity,del2016exact,arzamasovs2019full} in large deviations statistics in many-body systems. 
Emptiness formation probability (EFP) is probably the most iconic and widely studied measure of such large deviations. 
It plays a special role in the theory of Bethe Ansatz  \cite{bethe1931theorie} integrable models \cite{korepin1994correlation,korepin1997quantum,de2001six,boos2003emptiness} and is a test bed for development of non-perturbative techniques, such as the instanton calculus \cite{kleinert2009path}.   
The EFP, $P_{\mathrm EFP}(R)$, is the probability that no particles are found within the space interval $(-R,R)$ in the ground state of a one-dimensional (1D) many-body system with the average density $n$.
\begin{align}
										\label{eq:EFP}
P_{\mathrm EFP}(R) =  \prod_{i=1}^N \int_{|x_i| \geq R} dx_i\ |\Psi_g(x_1,x_2,...,x_N)|^2,
\end{align}
where $\Psi_g(x_1,x_2,...,x_N)$ is the normalized ground state wave function of N-particle system. Even in integrable models, where $\Psi_g$ is known via Bethe Ansatz, calculation of the multiple integral over the restricted interval is still a 
formidable task. A similar idea was first discuss in random matrix theory (RMT) \cite{mehta2004random}, where the probability that no eigenvalues fall within a certain interval of energy spectrum for different ensembles was studied \cite{dean2006large}.

For integrable systems, the problem is often formulated in terms of  spin-$1/2$  chains, where the EFP is defined as a probability of measuring $l$ consecutive spin to be  ``up''   in the ground state of the chain. Via  Jordan-Wigner transformation, such formulation is equivalent to the absence of  
quasiparticles on $l$ consecutive sites  \cite{shiroishi2001emptiness}. 
In these cases EFP is found to be expressed in terms of Fredholm determinants \cite{korepin1994correlation,kitanine2000correlation,shiroishi2001emptiness,kitanine2002emptiness,boos2006factorization}.   Even though EFP can be related to known mathematical constructions, extracting its asymptotic behavior is still extremely challenging. The exact answers are known so far only in a handful of isolated points in the parameter space \cite{shiroishi2001emptiness,Kitanine_2002}. 

This makes EFP an attractive playground for development of approximate asymptotic techniques. Most studies have been focusing on the regime $nR\gg 1$ (see, however, Ref.~\cite{bastianello2018exact}), where EFP is exponentially small,  $P_{\mathrm EFP}(R)\ll 1$. In this limit the problem may be studied within the semiclassical instanton approximation, where $\ln P_{\mathrm EFP}(R)$ is associated with (twice) the classical action along 
a certain dynamical trajectory of the Euler-Lagrange equations \cite{kleinert2009path}.  Such classical problem 
needs to be solved with the boundary conditions imposed on both ``past'' and ``future''  boundaries, which makes it 
not an easy task, neither analytically nor numerically. Similar structures are known in the theory of rare events in classical stochastic systems \cite{dykman1994large,elgart2004rare,krapivsky2012void,janas2016dynamical}.   

\begin{figure}[t]
    \centering
    \includegraphics[width=0.45\textwidth]{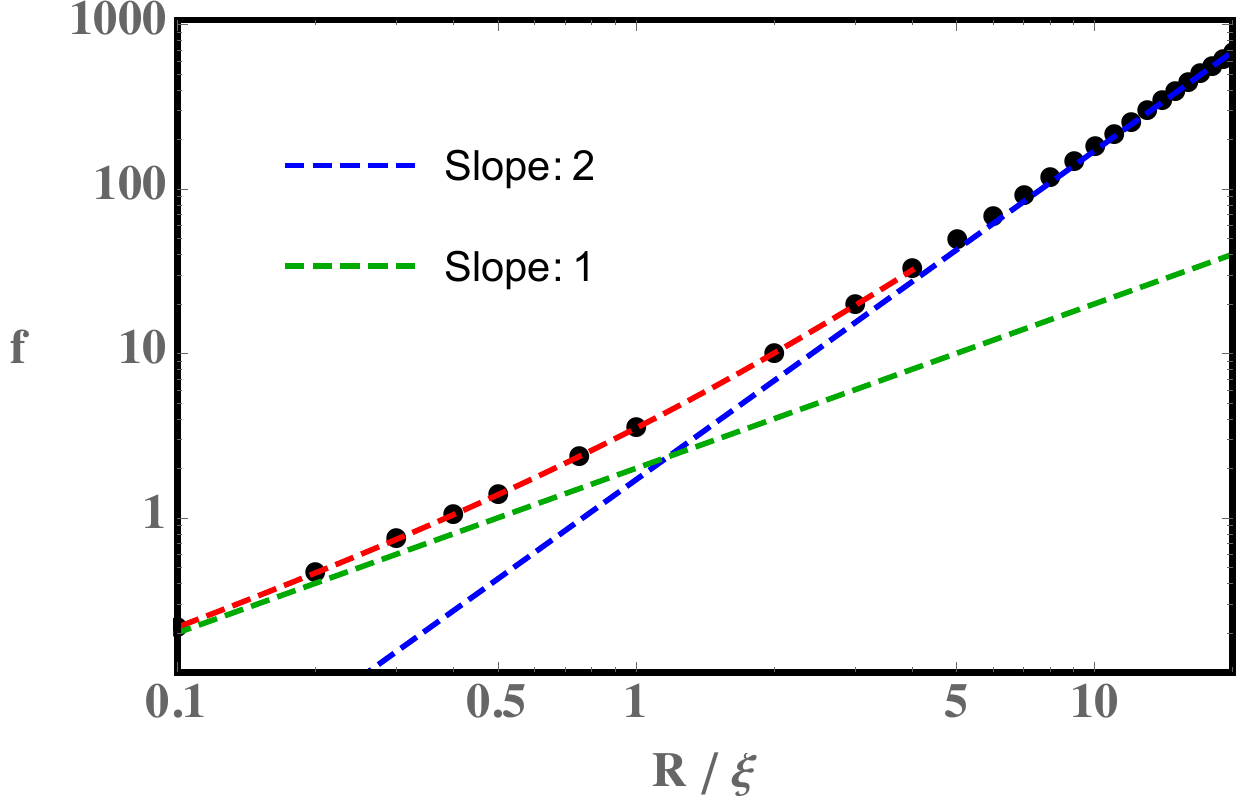}
    \caption{Function $f(R/\xi)$, Eq.~(\ref{EFP f(R/xi)}), for a weakly repulsive bosons in log-log scale. The numerical results show a crossover for the exponent of EFP  from linear (green dashed line) to quadratic (blue dashed line) and the red dashed line is fitted with first few points.}
    \label{FigCrossover}
\end{figure}

In this work we focus on EFP in the repulsive Lieb-Liniger (LL) model \cite{lieb1963exact},  of spinless bosons with the repulsive delta-potential in 1D. The ground (and excited) states of the model may be written through the Bethe ansatz \cite{lieb1963exact} and its thermodynamic characteristics are known exactly in terms of the microscopic parameters \cite{lieb1963exact}. In particular, one may find the sound velocity $v_s$ and thus define the healing (or correlation) length as 
 \begin{equation}
							\label{eq:xi}
\xi=(m v_s)^{-1}, 
\end{equation}
where $m$ is a mass of bosonic particles. In the limit of impenetrable interactions (the Tonks-Girardeau limit \cite{kulkarni2012hydrodynamics}), $\xi = (\pi n)^{-1}$, and the model is equivalent to the free fermions. Their (squared) ground state wave function  coincides with the joint probability distribution of eigenvalues in the circular unitary random matrix ensemble \cite{dyson1962statistical}. The exact answer for the free fermion EFP is thus known from RMT \cite{des1973asymptotic,dyson1976fredholm} 
\begin{align}
						\label{eq:free fermions} 
-\ln P_{\mathrm EFP}(R) = \frac{1}{2}\, \frac{R^2}{\xi^2} + \frac{1}{4}\ln(R/\xi) + O(1).
\end{align}
Within the instanton approach the leading term here was derived by A. Abanov \cite{abanov2004hydrodynamics} through a beautiful application of the complex valued functions theory.  The only treatment away from the Tonks-Girardeau  case, we are familiar with, is Ref.~\cite{korepin1995probability}, which conjectured EFP in the limit $\xi\ll R$, see our discussion below.        

Our particular focus here is on the opposite limit of the weakly interacting bosons. A defining feature of this regime is that the mean distance between the particles is much shorter than the correlation length, $n^{-1} \ll \xi$. As a result,  there is a  wide range $n^{-1}< R < \xi$, which was not previously discussed in the literature. 

Our main finding is that, through the entire range $n^{-1}\ll R$, the logarithm of EFP may be expressed as:  
\begin{align}
\label{EFP f(R/xi)}
-\ln P_{\mathrm EFP} = n\xi \times f\!\left(\frac R \xi \right),
\end{align} 
where $f(r)$ is a universal function, as long as $n^{-1}\ll \xi$, plotted in Fig.~\ref{FigCrossover}. Its asymptotic limits are:     
\begin{align}
& f(r) \approx 2.01(4)r + 1.50(4) r^2 +O(r^3); \quad\quad r\ll 1. 
										\label{EFP Poisson limit}
\end{align}
The leading term here is consistent with $P_{\mathrm EFP} \approx e^{-2Rn}$, which is the Poisson probability of finding the interval $2R$ empty of independent (i.e. non-interacting) randomly placed particles with   the mean density $n$. Indeed, the limit $R\ll \xi$ is reached in the non-interacting case (i.e. $\xi\to \infty$). The latter is characterized by the uniform ground state $|\Psi_g|^2=L^{-N}$, where $L=N/n$ is the system size. From Eq.~(\ref{eq:EFP}): $P_{\mathrm EFP}=\left(\frac{L-2R}{L}\right)^N\stackrel{N\to\infty}{\rightarrow}e^{-2Rn}$.

The other limit is:       
\begin{align}
&f(r) \approx 1.70(1)r^2 + 0.1(3)r + O(\ln r); \quad \quad r\gg 1. 
								\label{EFP Gaussian limit}
\end{align}
Now the leading term corresponds to the Gaussian EFP, $P_{\mathrm EFP} \approx \exp\{-1.7\,R^2 n/\xi\}$.
The Gaussian large-$R$ asymptotic of  the zero-temperature EFP may be argued on the very general ground \cite{abanov2004hydrodynamics}.
The specific coefficient, found here for the weakly interacting limit, is new.  It is at odds with the conjecture of Ref.~ \cite{korepin1995probability},   $-\ln P_{\mathrm EFP}= 4(R/\xi)^2$, which is parametrically  inconsistent with our scaling,  Eq.~(\ref{EFP f(R/xi)}).

The linear in $r$ term in Eq.~(\ref{EFP Gaussian limit}) is consistent with being zero. Indeed, in all cases with short-range 
interactions, where exact results are available \cite{shiroishi2001emptiness,Kitanine_2002}, such term is indeed absent. We believe that this is a generic feature of short-range interacting system and provide a perturbative argument to that effect in section \ref{III}.  Curiously, the Calogero-Sutherland model with the inverse square long-range interactions exhibits a non-zero $O(r)$ term (i.e. $\sim R$ term in the large $R$ asymptotic of  $-\ln P_{\mathrm {EFP}}$) \cite{sutherland1971quantum,mehta2004random}. Our numerical accuracy is not sufficient to establish a coefficient of $\ln R$ term in Eq.~(\ref{EFP Gaussian limit}).    

The paper is organized as follows. In Section \ref{II} we formulate an instanton approach for calculation of EFP for weakly interacting bosons. Numerical solution of corresponding Euler-Lagrange equations, discussion of the limiting cases and comparison with other works may be found in Section \ref{III}.   Appendix \ref{Appendix free fermions} is devoted to 
the free fermion limit as a test-drive of our numerical procedure.
 
\section{Instanton Calculus for Weakly Interacting Bosons}
\label{II} 
Here we adopt the hydrodynamic instanton approach to emptiness formation, developed by A. Abanov \cite{abanov2002probability,abanov2003emptiness,abanov2004hydrodynamics,franchini2005asymptotics,franchini2010emptiness}.
It is justified in the macroscopic emptiness regime, $n^{-1} \ll R$, where EFP is exponentially small. It is thus expected to be 
given by an optimal evolution trajectory in the space of the system's hydrodynamic degrees of freedom.  In our case the latter are the local particle density, $\rho(x,t)$, and the local current, $j(x,t)$. The two are rigidly related by the continuity equation, 
\begin{align}
\partial_t \rho + \partial_x j = 0.
\label{Continuity Equation}
\end{align}
The classical action, that yields proper hydrodynamic equations as its extremal conditions, is given by \cite{kulkarni2012hydrodynamics}  

\begin{align}
\label{NLSE action}
&S[\rho,j] = \int\!\!\! \int\! dxdt\, \Big[ \frac{mj^2}{2\rho} - V(\rho)\Big];\\
&V(\rho) = \frac{c}{2}(\rho - n)^2 + \frac{(\partial_x \rho)^2}{8m\rho}, 
\end{align}
where $m$ is the mass of a particle. The Lagrangian in Eq.~(\ref{NLSE action}) consists of the kinetic energy of the current along with the potential energy (equation of state) $V(\rho)$. For weakly interacting Bose liquid the latter is quadratic in density deviations from its equilibrium value, $n$, with the interaction parameter,  $c$. The correlation length is given by $\xi = 1/mv_s=(mnc)^{-1/2}$. It satisfies the weak interaction criterion, $n^{-1}\ll \xi$, as long as 
$\gamma \equiv mc/n \ll 1$. The potential energy also contains the so-called  quantum pressure \cite{kulkarni2012hydrodynamics,landau2013quantum} term, which reflects the tendency of the condensate to maintain the uniform density throughout the system (due to the gradient terms in the underlying quantum description). 

Variation of the action (\ref{NLSE action})  over  $\rho$ and $j$, under the continuity constraint, Eq.~(\ref{Continuity Equation}), yields classical Euler equation of the hydrodynamic flow (with the quantum pressure contribution) \cite{landau1987fluid}. Solutions of this equation do {\em not} lead to the formation of emptiness. The reason is that the emptiness is a large {\em quantum} fluctuation (similar to tunneling), which is located in a classically forbidden region of the phase space. 
The instanton approach is based on the realization that the quantum transition amplitude is given by the path integral $\int {\cal D}\rho {\cal D}j e^{iS[\rho,j]}\delta(\partial_t \rho + \partial_x j )$, with proper  boundary conditions. The integration contours over the field variables may be then deformed into the complex plane to pass through a classically forbidden  stationary configuration that reaches the required emptiness. The probability of such rare event is  $P\propto |e^{iS_{\mathrm inst}}|^2$, where the classical action along the instanton trajectory, $S_{\mathrm inst}$, acquires a (positive) imaginary part.

Before proceeding with the analytical continuation to the complex plane, it is convenient to pass from a Lagrangian formalism, Eq.~(\ref{Continuity Equation}), to the Hamiltonian one.  To this end we introduce a new auxiliary   
field $\partial_x \theta(x,t)$ and perform the  Hubbard-Stratonovich transformation for the kinetic energy term $\sim mj^2/(2\rho)$ in $e^{i S[\rho,j]}$. This brings  terms $ -\rho (\partial_x \theta)^2/(2m) + j\partial_x \theta$ to the action. One may then integrate by parts the last term (assuming periodic boundary conditions in the $x$ direction) and employ the continuity relation to  find:  
\begin{align}
S[\rho,\theta] = \int\!\!\!\int\! dxdt\,  \Big[\theta \partial_t \rho- \frac{\rho (\partial_x \theta)^2}{2m} - V(\rho)\Big],
\end{align}
where we neglected the factor $\sqrt{\det[\rho]}$ from the Hubbard-Stratonovich transformation, since it goes beyond the accuracy of the instanton approach. Notice that the fields $\rho$ and $\theta$ are not subject to any constraints and play the role of the canonical pair. 

We are now on the position to perform the analytical continuation. Following the standard treatment of tunneling, it is achieved by the Wick rotation to imaginary time $t \rightarrow -i\tau$. The resulting equations of motions may be solved with real $\rho$ and purely imaginary $\theta$ (the integration contour in $\theta$ is deformed to pass through an imaginary saddle point). It is convenient thus to  redefine $\theta \rightarrow i\theta$ such that the saddle point solutions for both $\rho$ and $\theta$ are real functions (in imaginary time), while the new $\theta$ integration runs along the imaginary axis. 
The corresponding Eucledian action acquires the Hamiltonain form
\begin{align}
\label{Euclidean action rho theta}
&S[\rho,\theta] = i\!\int\!\!\!\int\! dxd\tau\, [\theta\partial_\tau \rho - \mathcal{H}(\rho,\theta)];\\
&\mathcal{H}(\rho,\theta) = \frac{\rho (\partial_x \theta)^2}{2m} - \frac{c}{2}(\rho - n)^2 - \frac{(\partial_x \rho)^2}{8m\rho}.
\end{align}
Notice that the potential $V(\rho)$ enters the effective Hamiltonian, $\mathcal{H}(\rho,\theta)$, with the ``wrong'' sign, 
mirroring the inverted potential in the tunneling problem. 

The equations of motion, that follow from the action (\ref{Euclidean action rho theta}), are not the most convenient for the numerical solution. To facilitate the latter, we found useful to perform the canonical transformation $(\rho,\theta)\to (Q,P)$ to the new pair of the conjugated fields $Q(x,\tau)=\sqrt{\rho(x,\tau)}\, e^{-\theta(x,\tau)}$ and   $P(x,\tau)=\sqrt{\rho(x,\tau)}\,  e^{\theta(x,\tau)}$, or conversely $\rho=PQ$ and $\theta={1\over 2}\ln (P/Q)$.   Substituting these into Eq.~(\ref{Euclidean action rho theta}), one finds for the action
\begin{align}\nonumber 
&S[Q,P] = i\!\int\!\!\!\int\! dxd\tau\, [P\partial_\tau Q - \mathcal{H}(Q,P)]\\
&\qquad \quad\,\, + \frac{i}{2}\!\int\! dx\, PQ \ln\frac{P}{Q} \Big|_{\tau=\tau_i}^{\tau=\tau_f},
				\label{eq:actionPQ}\\
&\mathcal{H}(Q,P) = - \frac{\partial_x P \partial_x Q}{2m} - \frac{c(PQ-n)^2}{2},
\end{align}
where $\tau_{i(f)}$ are initial(final) times of the optimal trajectory, discussed below. 
  
Variables $Q,P$ may  be considered as an analytical continuation of the real-time degrees of freedom $Q\leftrightarrow \Psi$ and $P\leftrightarrow \Bar\Psi$. The first line of Eq.~(\ref{eq:actionPQ}) is nothing but the analytical continuation of the Gross-Pitaevskii (GP) action \cite{dalfovo1999theory}, $\sim |\partial_x\Psi|^2/2m+c(|\Psi|^2-n)^2/2$. However, would we start directly from the GP action, we would miss the boundary term, the second line in  Eq.~(\ref{eq:actionPQ}). This boundary term \footnote{The straight substitution results in the boundary term $i\int dx \rho(\theta-1/2) \big|_{\tau=\tau_i}^{\tau=\tau_f}$. There is however a global symmetry of shifting $\theta(x,t)\to \theta(x,t)+\bar\theta$ by a constant, associated with $P\to Pe^{\bar\theta}$ and    $Q\to Qe^{-\bar\theta}$, such that $\rho=QP$ is invariant. This allows to bring the boundary term to the form of  Eq.~(\ref{eq:actionPQ}).}, $i \int dx \rho\, \theta\big|_{\tau=\tau_i}^{\tau=\tau_f}$,  does not alter the equations of motion, but contributes to the instanton action. Its contribution appears to be of the paramount importance in the regime $n^{-1}<R<\xi$. To the best of our knowledge, it was first introduced in the context  of classical stochastic systems by Krapivsky, Meerson, and Sasorov~\cite{krapivsky2012void}, but was not discussed so far in the quantum context.    

It is convenient to pass to dimensionless coordinates and fields: $x \rightarrow \xi x$, $\tau \rightarrow \tau/(nc)$, $P \rightarrow \sqrt{n}P$, $Q \rightarrow \sqrt{n}Q$. In terms of them the Euclidean action takes the form 
\begin{align}
 \nonumber 
&S = i n\xi\left(\! \int\!\!\!\int\! dx d\tau\left[  P\partial_\tau Q + \frac{\partial_x P \partial_x Q}{2} + \frac{(PQ-1)^2}{2}\right]
\right. \\
\label{dimensionless Sb}
&\left. \qquad \qquad+ \frac{1}{2}\!\int\! dx\, PQ \ln\frac{P}{Q}\Big|_{\tau=\tau_f}^{\tau=\tau_i}\right).
\end{align}
The corresponding equations of motion acquire the universal parameter-free form: 
\begin{align}
\label{dimensionless Q}
&\partial_\tau Q = \frac{1}{2}\partial_x^2 Q - (PQ-1)Q,\\
\label{dimensionless P}
&\partial_\tau P = -\frac{1}{2}\partial_x^2P + (PQ-1)P.
\end{align}
These partial differential equations are known as Ablowitz-Kaup-Newell-Segur (AKNS) system \cite{ablowitz1973nonlinear}, which is integrable with the inverse scattering method. Remarkably, exactly these equations appear in the studies of rare events in Kardar-Parisi-Zhang classical stochastic equation \cite{meerson2016large,kamenev2016short,janas2016dynamical}.   

We can now specify the boundary conditions, appropriate for the emptiness formation problem. We are looking for a transition amplitude from a uniform state at a distant past, $\tau_i=-\infty$, to a state with the emptiness, i.e. zero density for 
$|x| < R$, at the observation time,  $\tau_f=0$. This leads to the conditions: $\rho(x,-\infty)=n$ and 
$\rho(|x| < R,0)=0$.  Outside of the interval $x\in (-R,R)$ at the observation time $\tau_f=0$, the density is not fixed and is to be integrated out in the boundary term $i \int dx \rho\, \theta\big|^{\tau=\tau_f}$ . This fixes $\theta(|x|>R,0)=0$. In terms of the dimensionless coordinates and fields $Q,P$, these read as:  
\begin{align}
\label{BC2}
&PQ(x,-\infty) = 1; \\
					\label{BC1}
&P(x, 0) = \Big\{\begin{array}{ll}
     0,& \,\,\, |x| <R/\xi, \\
     Q(x,0),&\,\,\, |x| > R/\xi.
\end{array}
\end{align}
The zero density constraint within the emptiness interval $\rho=QP=0$, may be enforced by either $P=0$, or $Q=0$. This choice is arbitrary, since $Q$ and $P$ are interchangeable by a canonical transformation.

The program now is as follows: one needs to solve the stationary field equations (\ref{dimensionless Q}) and (\ref{dimensionless P}), subject to the boundary conditions (\ref{BC2}) and (\ref{BC1}). The resulting instanton trajectory is to be substituted into the action (\ref{dimensionless Sb}) (including the boundary term (!)), resulting in the instanton action 
$S_{\mathrm inst}(R)$. The semiclassical transition amplitude is then given $e^{iS_{\mathrm inst}(R)}$, resulting finally in the EFP of the form  
\begin{align}
\label{EFP Sopt}
-\ln P_{\mathrm EFP}(R) = 2\Im S_{\mathrm inst}(R).
\end{align}
One notices then that the Eqs.~(\ref{dimensionless Q}), (\ref{dimensionless P}) are free from any parameters, while the boundary conditions (\ref{BC2}), (\ref{BC1}) depend on the single parameter, $R/\xi$. The form of the action (\ref{dimensionless Sb}) immediately implies  the result, Eq.~(\ref{EFP f(R/xi)}), where $f(R/\xi)$ is twice the value of the double integral plus the boundary term, within the large round brackets on the right hand side of Eq.~(\ref{dimensionless Sb}), evaluated along the optimal trajectory.

\section{Results and Discussion}
\label{III}

The equations of motion (\ref{dimensionless Q}), (\ref{dimensionless P}) are of the AKNS type and thus are, in principle,   integrable. However, the boundary conditions (\ref{BC2}), (\ref{BC1}) are {\em not} the initial value problem, which could be treated with the inverse scattering approach. Although a lot is known about solutions of Eqs.~(\ref{dimensionless Q}), (\ref{dimensionless P}) (see, eg., discussion of their multi-soliton configurations in Ref.~\cite{janas2016dynamical})), we were not able to find their analytical treatment, suitable for EFP setup, formulated above. We thus resorted to a numerical approach.

We use Chernykh-Stepanov  algorithm \cite{chernykh2001large,elgart2004rare} to solve  equations of motion iteratively. The algorithm takes the advantage of the diffusive character of Eq.~(\ref{dimensionless Q}) in the {\em forward} time and Eq.~(\ref{dimensionless P}) in the {\em backward} time. The two equations are successively evolved 
$Q$-forward, followed by $P$-backward in time to converge to  desired solutions. The diffusive character of the equations provides  stability for such iteration scheme, making the $(Q,P)$ variables advantageous over the $(\rho,\theta)$ pair. 
The results are still presented in terms of the more physically intuitive $(\rho,\theta)$ degrees of freedom.

At the initial backward-propagating step, we put $Q(x,\tau) = 1$ and $P(x,0) = \theta(|x|-R)$, here $\theta(x)$ is the Heaviside step function. Then $P(x,\tau)$ is determined from backward evolution of  Eq.~(\ref{dimensionless P}) up to a large negative time $\tau = -T$. Next we update the initial condition for $Q$ from $Q(x,-T)P(x,-T)=1$, cf. Eq.~(\ref{BC2}), and  evolve  Eq.~($\ref{dimensionless Q}$)  forward in time up to $\tau=0$, with $P(x,\tau)$ found in the first step. 
This way we obtain new $Q(x,\tau)$, which we use to update initial conditions for $P$ at $\tau=0$, according to Eq.~(\ref{BC1})  and evolve $P$ backward in time again, {\em etc}. We then evaluate the action (\ref{dimensionless Sb}), 
and check that its value does not depend on the choice of the large negative initial time, $-T$.   

\begin{figure}[h!]
    \centering
    \includegraphics[width=0.45\textwidth]{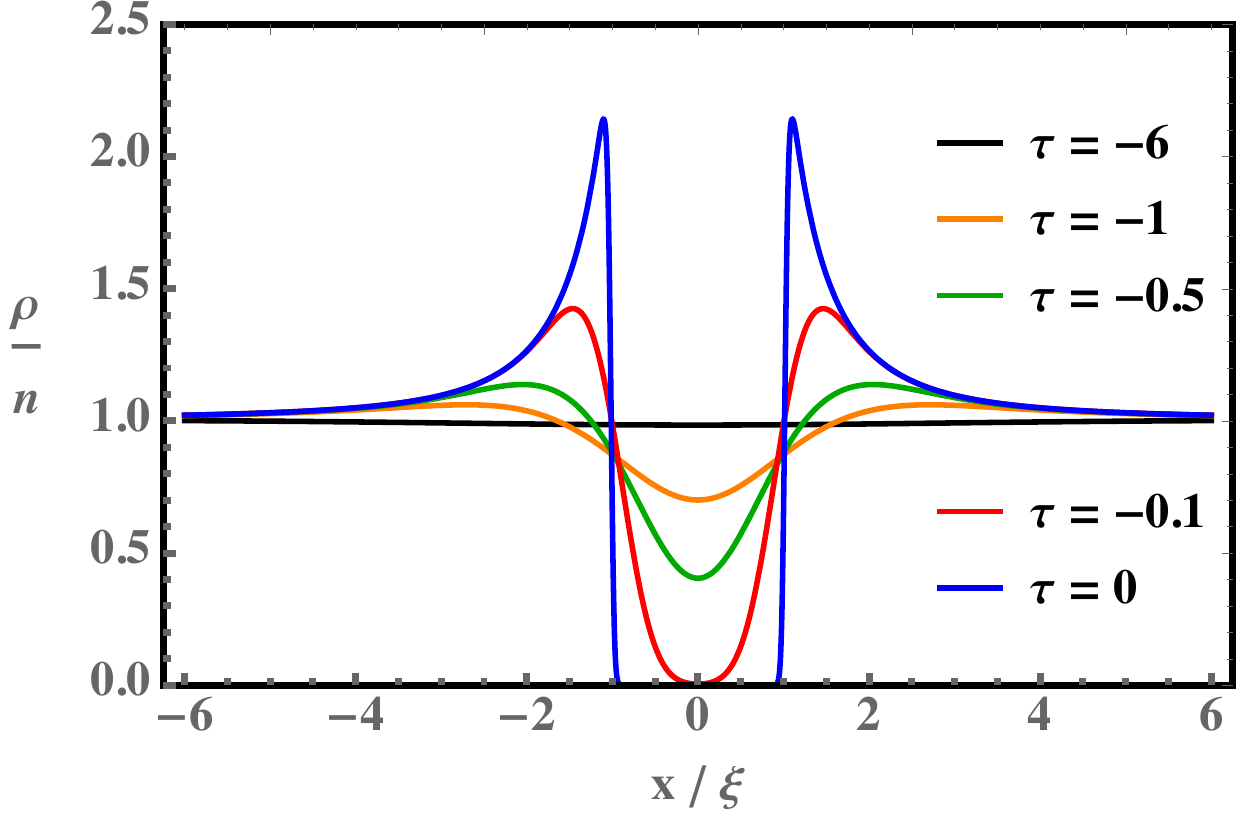}
    \includegraphics[width=0.45\textwidth]{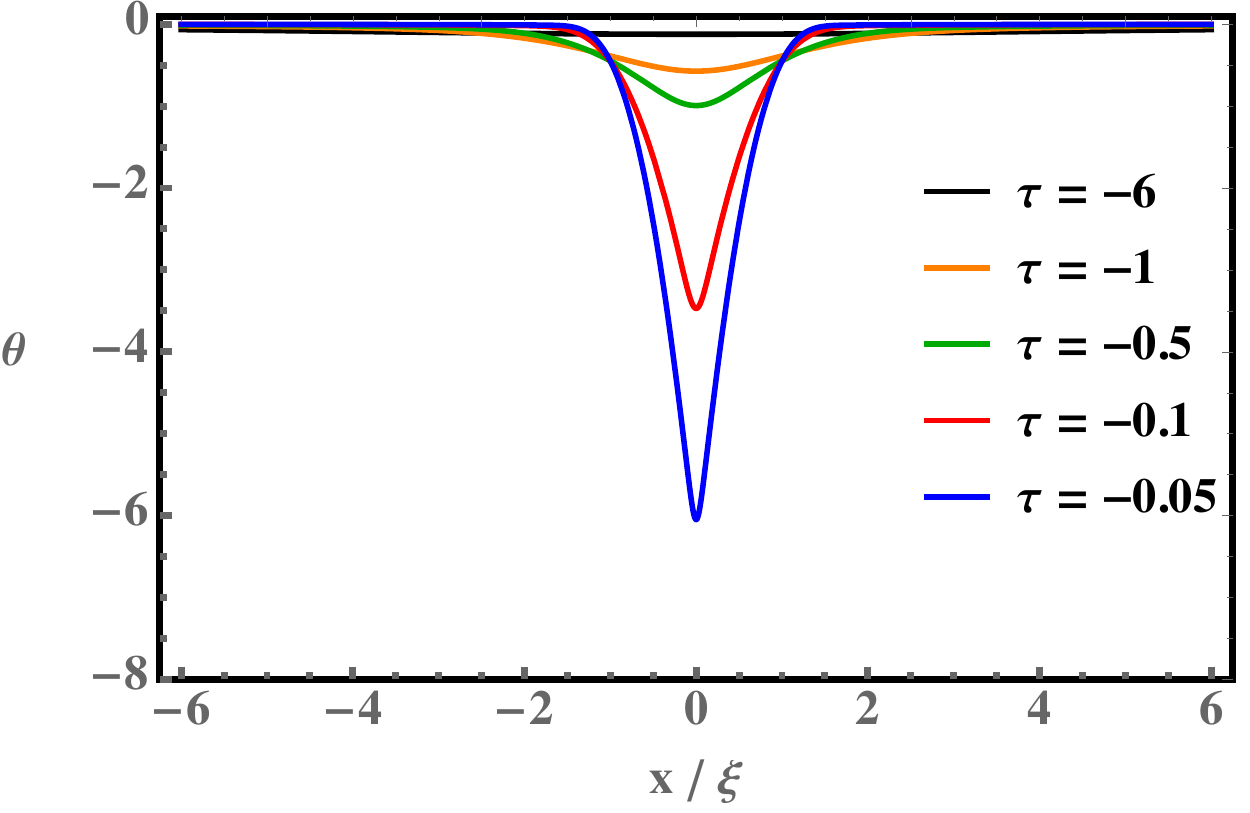}
    \includegraphics[width=0.45\textwidth]{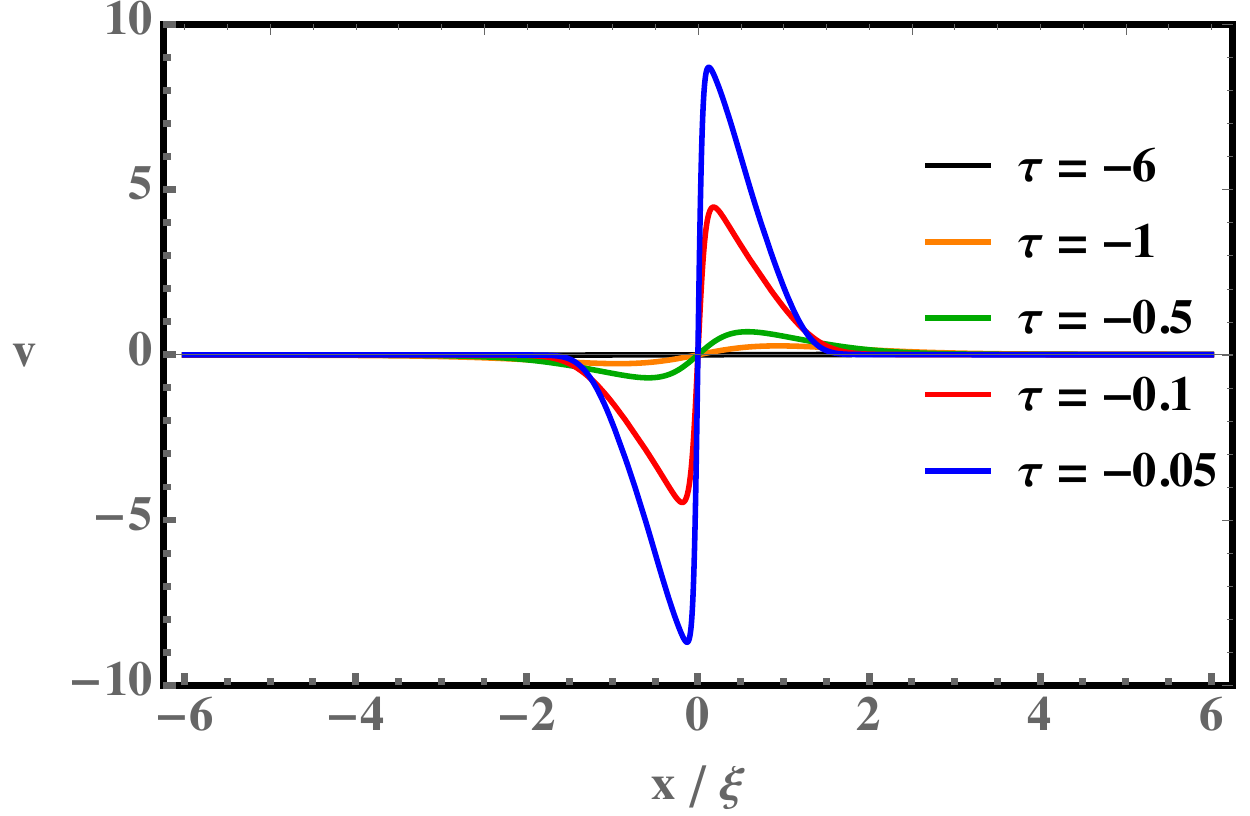}
    \caption{Time evolution of the density $\rho(x,\tau)$, imaginary phase $\theta(x,\tau)$ and velocity $v(x,\tau)$ for weakly interacting bosons with $R/\xi = 1$. The density evolves from the uniform value, $\rho =n$, at large negative $\tau$  towards the emptiness of size $2R$ at $\tau = 0$. At $\tau \to 0$ the phase tends to the negative infinity for $|x|<R$ and thus is not shown. The velocity v is related to spatial gradient of phase $v = \partial_x \theta/m.$ }
    \label{FigInstanton-r1}
\end{figure}

\begin{figure}[h!]
    \centering
    \includegraphics[width=0.45\textwidth]{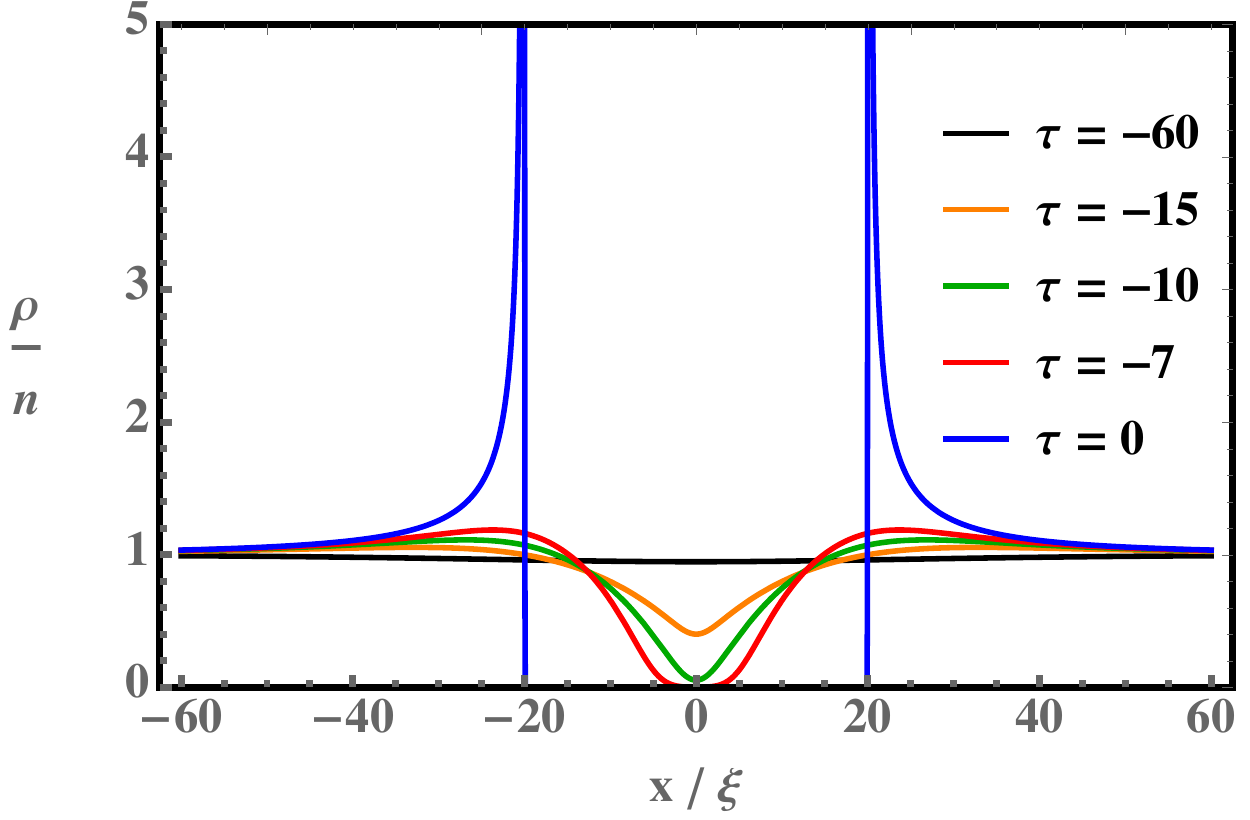}
    \includegraphics[width=0.45\textwidth]{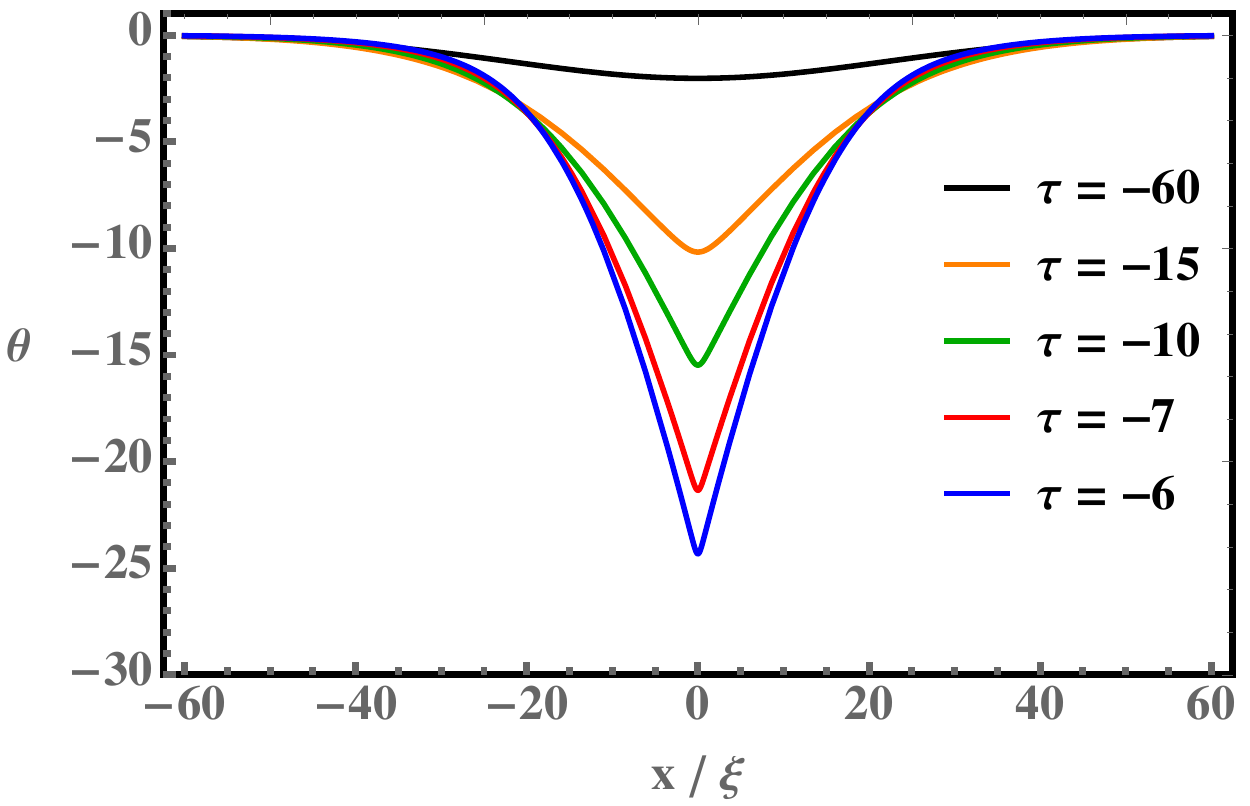}
    \includegraphics[width=0.45\textwidth]{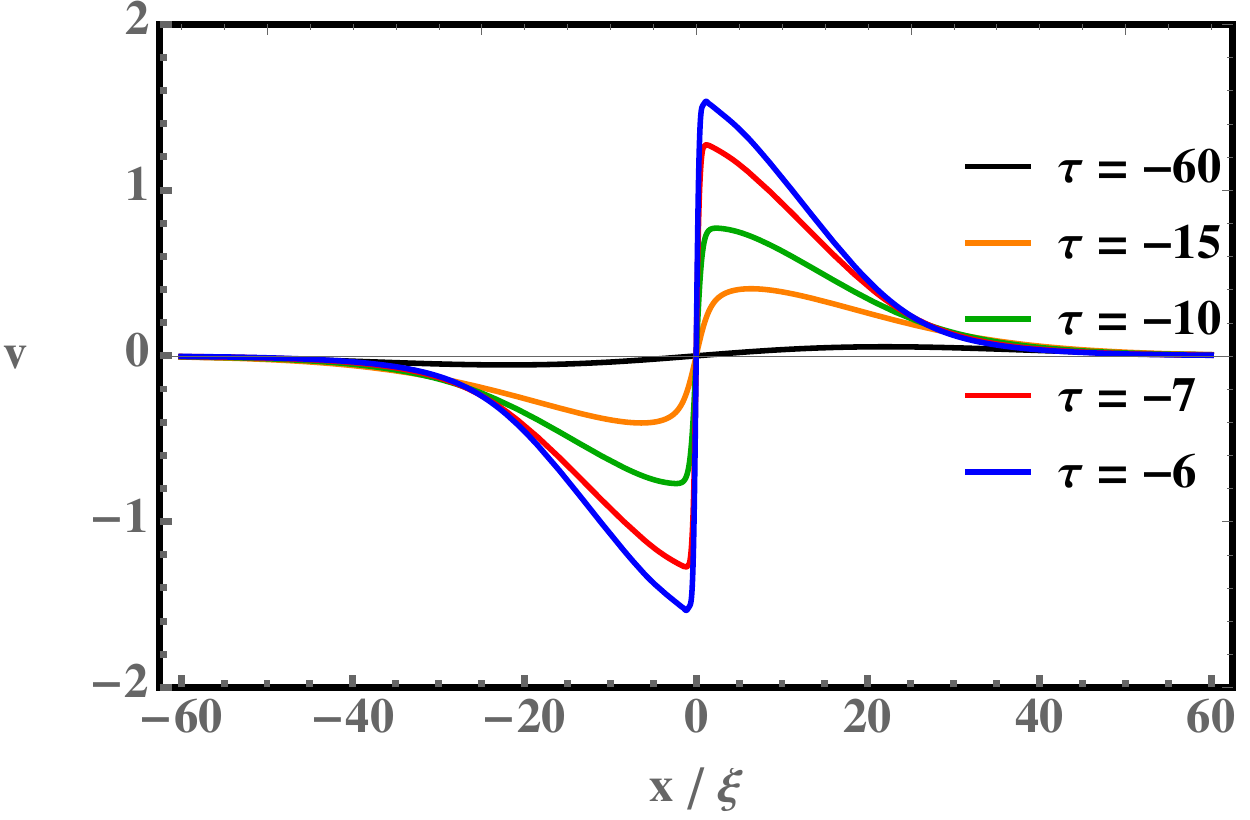}
    \caption{Same as Fig.~\ref{FigInstanton-r1} for $R/\xi = 20$.
    }
    \label{FigInstanton-r20}
\end{figure}

The evolution of density and (imaginary) phase are shown in Figs.~\ref{FigInstanton-r1} and \ref{FigInstanton-r20} for $R/\xi=1$ and $20$.    
The corresponding $f(R/\xi)$ is presented in Fig.~\ref{FigCrossover}. Its numerical fits in the regimes $R/\xi\ll 1$ and $R/\xi\gg 1$ are summarized in Eqs.~(\ref{EFP Poisson limit}) and (\ref{EFP Gaussian limit}) correspondingly. 

In  $R/\xi \ll 1$ limit the system is approaching the non-interacting one. Indeed, $\xi\to \infty$ is equivalent to $c\to 0$ limit.  
In this case the stationary equations for $Q$ and $P$ become pure diffusion and anti-diffusion, while the dynamical part of the action (\ref{dimensionless Sb}) is $\int\!\!\!\int\! dx d\tau P[\partial_\tau Q-{1\over 2}\partial_x^2 Q]$, which is nullified on the equation of motion.   The only contribution to the action is thus the boundary term $\frac{1}{2}\int\! dx PQ \ln\frac{P}{Q}\big|_{\tau_i=-\infty}\!\!=\int\! dx \rho\,\theta\big|_{\tau_i=-\infty}\!\!=- \int\! dx\,\theta(x,-\infty)$ (the final time, $\tau_f=0$, does not contribute either in view of Eq.~(\ref{BC1})). It numerical evaluation gives $1.005(20)$ in the limit $R\ll \xi$.

In the opposite $R/\xi \gg 1$, it is useful to look at the action (\ref{Euclidean action rho theta}) and rescale variables in an  alterantive way: $x \rightarrow R x$, $\tau \rightarrow Rm\xi \tau$, $\rho \rightarrow n\rho$ and $\theta \rightarrow (R/\xi)\theta$. The Euclidean action takes the form 
\begin{align}
\label{Hydro action}
S =& i n\xi\!\int\!\!\! \!\int\!dx d\tau\ \Big[\frac{R^2}{\xi^2} \left(\theta \partial_\tau \rho - \frac{\rho (\partial_x \theta)^2}{2} + \frac{(\rho-1)^2}{2} \right)    \nonumber \\
+& \frac{(\partial_x \rho)^2}{8\rho}\Big].
\end{align}
The first line here is the leading term, $\propto (R/\xi)^2$, which is given by the hydrodynamic action without quantum pressure. It corresponds to the leading Gaussian term in EFP, Eq.~(\ref{EFP Gaussian limit}). One notices the absence of the linear, in $R/\xi$, term, consistent with our numerical finding. The quantum pressure correction (the second line in Eq.~(\ref{Hydro action}), seems to be of the order $O(1)$. This may be misleading, since an attempt to treat the quantum pressure as a perturbative correction, seems to lead to logarithmically divergent integrals. This is probably the reason why the leading correction to the Gaussian result is of the order of $O(\ln R/\xi)$, cf. Eq.~(\ref{EFP Gaussian limit}). However this type of terms exceeds the accuracy of the instanton approximation. 

The message from Eq.~(\ref{Hydro action}) is that the Gaussian part of EFP in the limit $R\gg\xi$, can be found without the quantum pressure. This is in agreement with the success of such hydrodynamic theory  \cite{abanov2004hydrodynamics}  to obtain exact results vis-a-vis the Gaussian limit. Most notable case is the free fermion 
Tonks-Girardeau limit, cf. Eq.~(\ref{eq:free fermions}). We have numerically explored this known limit, see Appendix \ref{Appendix free fermions}, as a test-drive of our numerical procedure. We have found coefficient $0.501(2)$, which should be compared with $1/2$ in Eq.~(\ref{eq:free fermions}) - this provides some support to the accuracy of our results.

We conclude with a brief comparison with some previously published results on EFP.  
The only analytic work,  we know of, on EFP in 1D interacting boson model is a conjecture by Its, Korepin and Waldron \cite{korepin1995probability}. In the weakly interacting limit, the leading term at large $R$ is claimed to be 
$ -\ln P_{\mathrm EFP} = 4(R/\xi)^2$. This is in a parametric disagreement with our main result (\ref{EFP f(R/xi)}).  On the other hand, calculations based on the bosonisation procedure \cite{abanov2004hydrodynamics}  are in a parametric agreement with Eq.~(\ref{EFP f(R/xi)}).  Bosonization only allows for a treatment of a small suppression of density, rather than the emptiness. If one arbitrarily takes such ``small'' supression all the way to zero density, its probability is consistent with Eq.~(\ref{EFP f(R/xi)}).   

There is also a number of results on EFP in  antiferromagnetic spin-1/2 XXZ chain with the Hamiltonian
\begin{align}
H = \sum_{j = -\infty}^{\infty}\! \left[ S_j^x S_{j+1}^x + S_j^y S_{j+1}^y + \Delta S_j^z S_{j+1}^z\right],
\end{align}
where $\Delta$ is the anisotropy in the $z$-direction. It is proposed in Ref.~\cite{korepin2003asymptotic} that EFP in the gapless regime, $-1 < \Delta \leq 1$, is
\begin{align}
\label{XXZ astptotics}
-\ln P_{\mathrm EFP} \sim A l^2 + B \ln l,
\end{align}
where $A$ and $B$ are constants depending on $\Delta$ and $l$ is the number of consecutive spin-polarized 
sites. An explicit expression for the coefficient $A$ was found to be
\begin{align}
\label{XXZ chain leading term coefficient}
A = \ln\Big[\frac{\Gamma^2(1/4)}{\pi\sqrt{2\pi}}\Big] - \int_0^{\infty} \frac{dt}{t} \frac{\sinh^2(t\nu) {\rm e}^{-t}}{\cosh(2t\nu)\sinh(t)},
\end{align}
where parameter $\nu$ is defined through $\cos(\pi \nu)=\Delta$. The correspondence with the weakly interacting bosons may be established for $\Delta\gtrsim -1$, where the Luttinger parameter $K=1/[2(1-\nu)]\gg 1$. Defining the correlation length (in lattice units) as $\xi=K/(\pi n)$, where the corresponding bosonic density is $n=1/2$ in the lattice units, one finds from Eq.~(\ref{XXZ chain leading term coefficient}) for the leading term of EFP  in $\xi\gg 1$ limit: 
\begin{align}
-\ln P_{\mathrm EFP} = \frac{n}{2\xi}\, l^2,
\end{align}
which is in parametric agreement with our result (\ref{EFP f(R/xi)}). 

To conclude, we have developed the instanton approach that is capable to describe  a complete crossover of EFP from the Poisson to the Gaussian regime in the wide range of parameters, $n^{-1} < R <\xi$, available in weakly interacting bosonic 1D systems. Such systems are now routinely realized in cold atom experiments, where EFP may be measured.

\section{Acknowledgments} 

We are indebted to A. Abanov, D. Gangardt and B. Meerson for valuable discussions. This work was supported by NSF grant DMR-1608238. 

\appendix
\section{Free Fermions Limit}
\label{Appendix free fermions}

\begin{figure}[h!]
    \centering
    \includegraphics[width=0.45\textwidth]{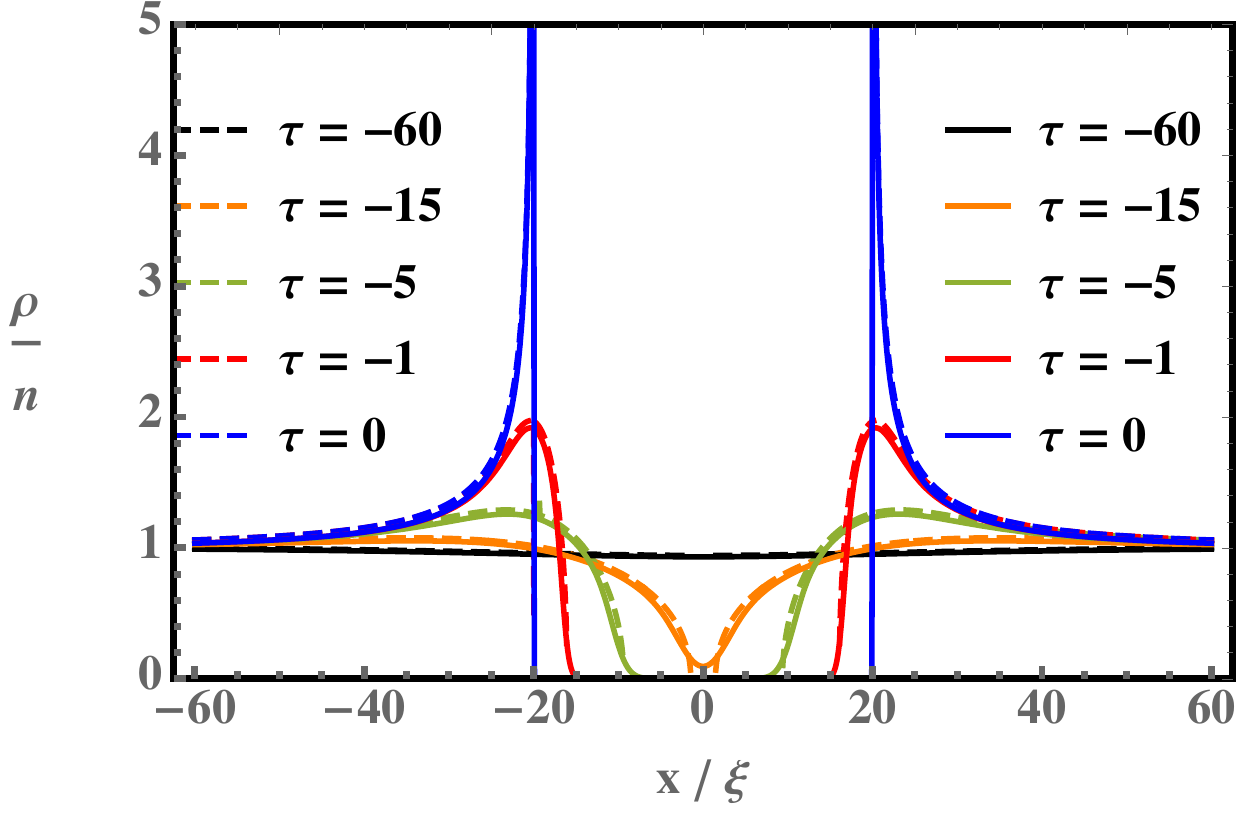}
    \includegraphics[width=0.45\textwidth]{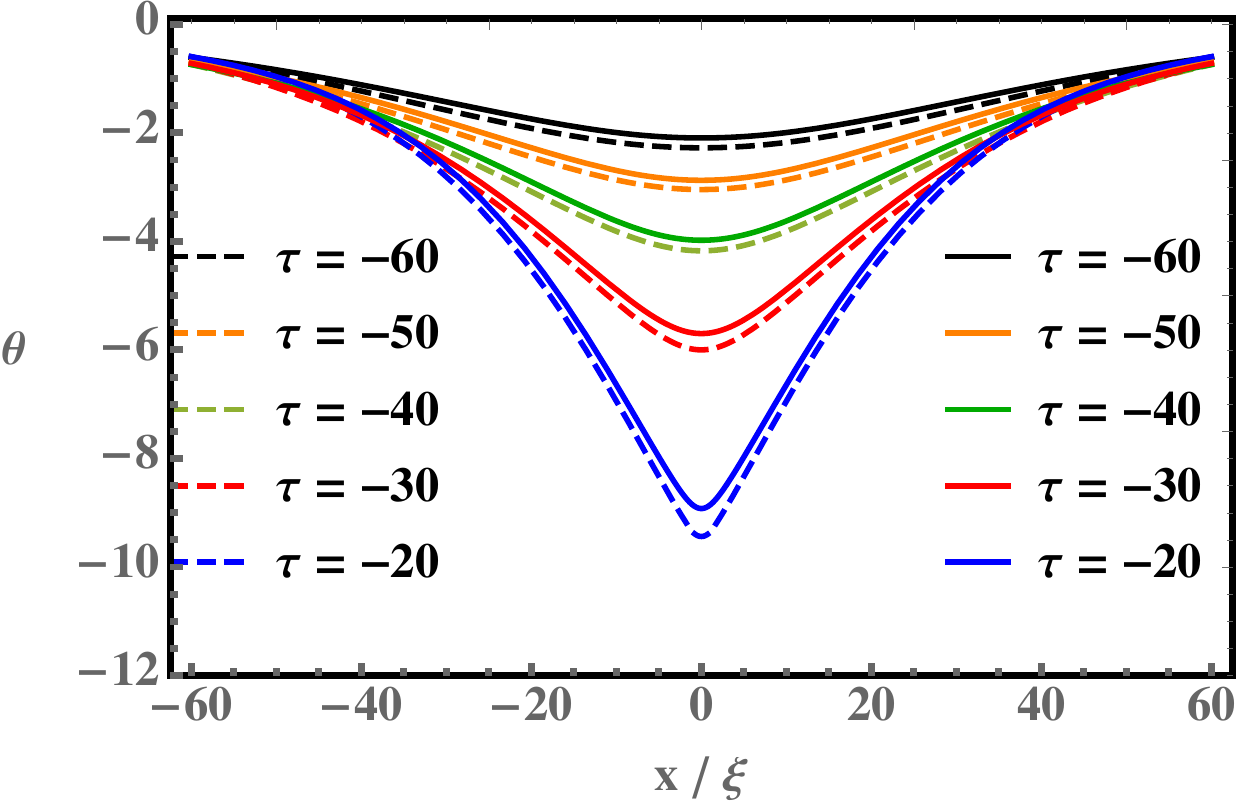}
    \includegraphics[width=0.45\textwidth]{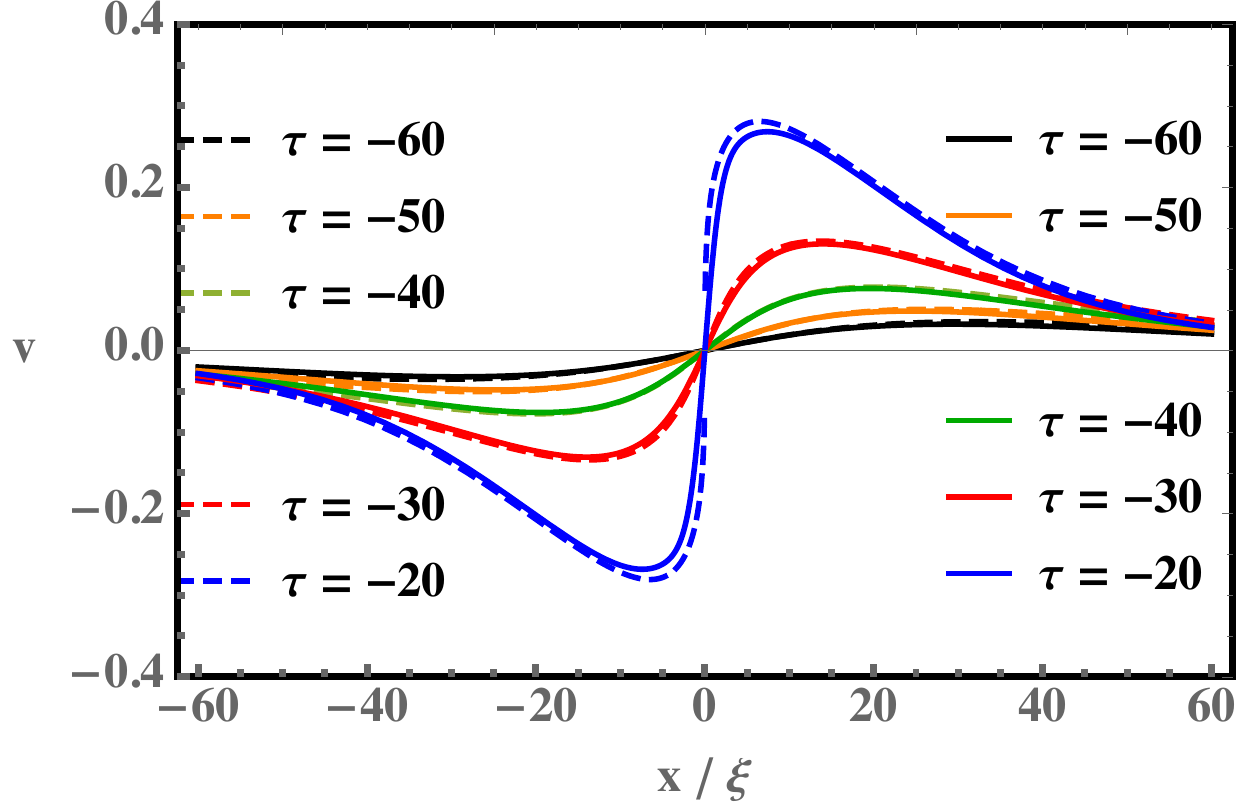}
    \caption{The upper panel is the time evolution of the density $\rho$ for free fermions with $R/\xi = 20$. The middle and lower one is the time evolution of phase $\theta$ and velocity $v$. The solid lines are numerical solutions of Eqs.~(\ref{modified dimensionless Q}), (\ref{modified dimensionless P}), using the algorithm outlined in Section \ref{III}, and dashed lines are the analytical solutions of Ref.~\cite{abanov2004hydrodynamics}. }
    \label{FigInstantonFermion}
\end{figure}

\begin{figure}[h!]
    \centering
    \includegraphics[width=0.45\textwidth]{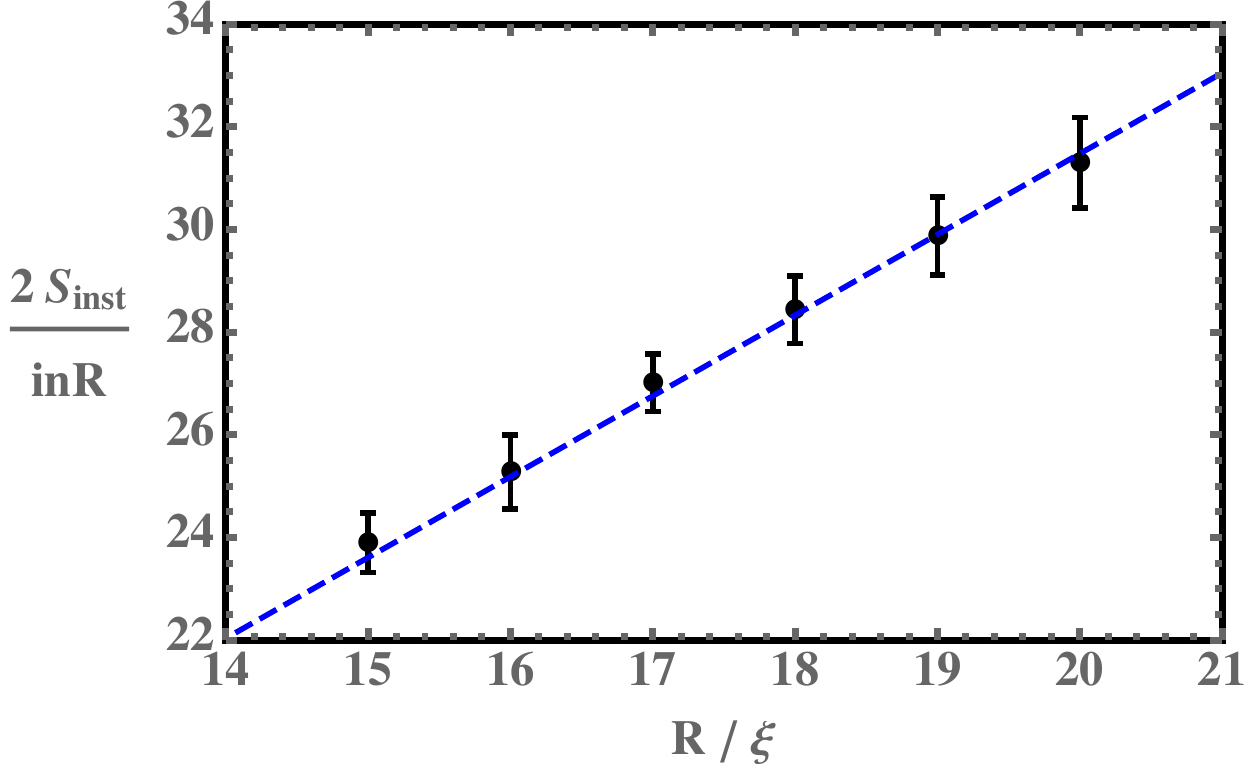}
    \caption{$2\Im S_{\mathrm inst}/nR$ vs. $R/\xi$ for free fermions. The blue dashed line is a fit $0.501\pi R/\xi$.}
    \label{FigSoptFermion}
\end{figure}

In the free fermion limit the hydrodynamic potential is given by 
\begin{equation}
V(\rho)= 2\int_{\pi n}^{\pi\rho} \frac{dk}{2\pi} \frac{k^2}{2m}-\mu(\rho-n)= \frac{\pi^2(\rho^3-3n^2\rho+2n^3)}{6m},
\end{equation}
where $\mu=(\pi n)^2/2m$ is the chemical potential. We substitute it in the hydrodynamic action (\ref{Euclidean action rho theta}) 
to find  
\begin{align}
\label{Hydro action fermion}
S = i\!\int\!\!\!\int\! dx d\tau \Big[\theta \partial_\tau \rho - \frac{\rho (\partial_x \theta)^2}{2} + V(\rho) + \frac{(\partial_x \rho)^2}{8\rho}\Big],
\end{align}
where we kept the quantum pressure term from the weakly interacting case, since,  as explained in Section \ref{III}, it does not contribute in the large $R$ limit anyways. We now proceed to the $Q,P$ variables as above and then make them dimensionless, using $\xi = 1/\pi n$ appropriate for the free fermions. The resulting equations of motion are
\begin{align}
\label{modified dimensionless Q}
&\partial_\tau Q = \frac{1}{2}\partial_x^2 Q - \frac{1}{2}(P^2Q^2-1)Q,\\
\label{modified dimensionless P}
&\partial_\tau P = -\frac{1}{2}\partial_x^2P + \frac{1}{2}(P^2Q^2-1)P,
\end{align}
with the same boundary condition ($\ref{BC2}$), ($\ref{BC1}$) and the modified action
\begin{align}
S&= in\xi\left( \int\!\!\!\int \! dx d\tau \Big[P\partial_\tau Q + \frac{\partial_x P \partial_x Q}{2}  \right., \\
&  \left. + \frac{(P^3Q^3-3PQ+2)^2}{6}\Big]  + \frac{1}{2}\int dx\ PQ \ln\frac{P}{Q}\Big|_{\tau=-\infty}^{\tau=0}\right).
\nonumber
\end{align}
The instanton solution is shown in Fig.~\ref{FigInstantonFermion}, where we compare it to the analytical solution (without quantum pressure) of Ref.~\cite{abanov2004hydrodynamics}.  The corresponding optimal action is  shown in 
Fig.~\ref{FigSoptFermion}. Its best fit is given 
\begin{align}
-\ln P_{\mathrm EFP} = 0.501(2)(R/\xi)^2 + O(\ln R/\xi),
\end{align}
where we used relation $\xi = 1/\pi n$. This is in a very good agreement with exact result for the free fermions, Eq.~(\ref{eq:free fermions}), \cite{des1973asymptotic,dyson1976fredholm}.

\newpage
\bibliography{draft.bib}

\end{document}